\documentstyle[twocolumn,epsf,aps]{revtex}
\draft
\begin{document}
\title{Generalized contact process on random environments}

\author{Gy\"orgy Szab\'o$^1$, Hajnalka Gergely$^2$, and
Be\'ata Oborny$^2$}
\address
{$^1$Research Institute for Technical Physics and Materials Science \\
P.O.Box 49, H-1525 Budapest, Hungary \\
$^2$Department of Plant Taxonomy and Ecology, E\"otv\"os University \\
H-1117 Budapest, P\'azm\'any P\'eter s\'et\'any 1/C, Hungary}

\address{\em \today}

\address {
\centering {
\medskip \em
\begin{minipage}{15.4cm}
{}\qquad Spreading from a seed is studied by Monte Carlo
simulation on a square lattice with two types of sites affecting the
rates of birth and death. These systems exhibit a critical transition
between survival and extinction. For time-dependent background, this
transition is equivalent to those found in homogeneous systems
(i.e. to directed percolation). For frozen backgrounds, the appearance
of Griffiths phase prevents the accurate analysis of this transition.
For long times in the subcritical region, spreading remains localized
in compact (rather than ramified) patches, and the average number
of occupied sites increases logarithmically in the surviving trials.
\end{minipage}
}}

\maketitle

\narrowtext

\section{INTRODUCTION}
\label{sec:int}

Contact process (CP) was introduced by Harris \cite{h74}
to describe competition between death and birth events
for a spatially distributed species. In this model, the
organisms localized at the sites of a lattice can die with a
death rate or can create additional offspring in one of their
(empty) neighboring sites. When increasing the ratio of birth to
death rates, this system exhibits a critical phase transition
from empty to the active state as the system size approaches
infinity. This transition belongs to the so-called directed
percolation (DP) universality class \cite{j81,g82}. Similar
transitions have been observed for some other, related models 
(for recent reviews see the Refs. \cite{md99,h00}).

The above mentioned CPs applied homogeneous backgrounds.
Numerical investigations on two-dimensional random environments
have been restricted to diluted lattices 
\cite{n86,n88,md96,dm98,ttn}. Noest \cite{n86} have shown
that, although the
transition remains continuous, the extinction process is
modified drastically on the randomly diluted lattice above
the percolation threshold. Recently Dickman and Moreira
\cite{dm98} have reported that the scaling (a fundamental
feature of critical transitions) is violated in this case.
According to a field-theoretic analysis, similar result is
predicted by Janssen \cite{j97}.

Below the percolation threshold on a diluted lattice, the
available sites form isolated (finite) clusters. In any 
particular cluster the species is expected to die out within
finite time and the cluster remains empty afterwards. A simple
mathematical description suggested by Noest \cite{n88} indicates
that the average concentration of species vanishes algebrically
with time and this phenomenon is analogous to the relaxation
observed for the ``Griffiths phase'' in disordered spin
models \cite{g69}.

Field-theoretic arguments \cite{h00} support that the temporally
quenched disorder itself, as well as the spatially quenched disorder,
disturbs crucially the DP transition. However, DP transition is
expected on large scales for those random environments where the
spatio-temporal disorder is uncorrelated. For two-dimensional systems
this expectation is supported only by a small number of numerical
evidences. For example, in an evolutionary game, extinction of one
of three competing strategies represents a DP transition on
a background whose time dependence is maintained by the competition
of the remaining two strategies \cite{sasd}. 

In the present paper, we study a generalized CP on a square
lattice with two types of sites that provide different conditions
for survival and reproduction for a hypothetical species.
Our analysis is not restricted to frozen backgrounds.
We give numerical evidence that the extinction process becomes
analogous to DP transition if the random environment involves
uncorrelated time dependence. Using numerical
simulations, we investigate the main characteristics of spreading
when initially there is only a single individual in the system.
This method was suggested by Grassberger and de la Torre
\cite{gdt} to study DP transition (for homogeneous background)
in the close vicinity of the critical point. Moreira and Dickman 
\cite{md96} have demonstrated that this technique is also
efficient for the investigation of inhomogeneous systems. Now
we have adopted this method to study the above-mentioned generalized
CP in random environments. The results confirm the theoretical
expectations \cite{md99,h00,j97,m90} as well as the previous
observations based on MC simulations \cite{n86,n88,md96,dm98,ttn}.
We have also studied some geometrical features of the patches
formed by occupied sites in the subcritical region.

\section{THE MODEL} 
\label{sec:model}

Space is represented by a square lattice. Each lattice site, ${\bf r}
=(x,y)$ ($x$ and $y$ are integers), is assumed to represent a microhabit
for one individual of a species. Quality of a site, $g({\bf r})$,
determines the probabilities of survival and reproduction of the
inhabitant individual. We assume two microhabitat types: good and bad
(where $g({\bf r})=1$ or $0$, respectively). Site qualities are
distributed randomly in space. Initially, each site is chosen to be
good (bad) with a probability $P$ ($1-P$). Distribution of individuals
over this random background is described by a state variable
$\sigma({\bf r})$ that is 1 for occupied and 0 for empty state. Time
evolution of the system is governed by three elementary processes
affecting the values of $\sigma ({\bf r})$ and $g({\bf r})$.
For an occupied site, extinction (death) can occur with a death rate
$d[g({\bf r})]$. For an empty site, colonization (birth) can take
place with a birth rate $b[g({\bf r})]$, provided that a randomly
chosen nearest-neighbor site is occupied. It is supposed that good
sites provide better conditions for living, that is $b(1)>b(0)$ and
$d(1)<d(0)$. Furthermore, redrawing the value of $g({\bf r})$
we change the environment with a rate $f$. The new state is
independent of the actual values of $g({\bf r})$ and
$\sigma({\bf r})$. This occasional modification results
in stationary distribution of good or bad sites over time, with 
probabilities $P$ and $1-P$, respectively, and leaves their distribution
uncorrelated.

In the special case when every site is good ($P=1$), the present
model is equivalent to the well investigated contact process on
homogeneous background \cite{md99,h00}.
Previous studies have shown that in this case, the species dies out
if $\lambda = b(1)/d(1) > \lambda_c = 1.6488(1)$. Concentration of
population vanishes algebrically in the active stationary state,
that is $c \propto (\lambda - \lambda_c)^{\beta}$ where
$\beta = 0.575(3)$ if $\lambda \to \lambda_c$ \cite{md99,h00,ad,bfm}.
This critical transition is accompaned with diverging
fluctuations and correlation length.

The conditions of CP on diluted lattice can be reproduced by
allowing the species to stay only in good sites ($b(0)=0$ and
$d(0)=\infty$) for $P < 1$ and $f=0$.

A mean-field approximation for frozen backgrounds ($f=0$)
predicts different concentration of species on good and
bad sites: $c_1$ and $c_0$, respectively. Neglecting the 
details of a straightforward calculation, the concentrations obey
the following expressions:
\begin{equation}
c_1={\lambda_1 c \over 1 + \lambda_1 c} \;\;\; \mbox{ and } \;\;\;
c_0={\lambda_0 c \over 1 + \lambda_0 c} \,\, , 
\label{eq:mf1}
\end{equation}
where $\lambda_{\alpha}=b(\alpha)/d(\alpha)$ ($\alpha=0$ or 1),
and the average concentration, $c=P c_1 + (1-P)c_0$ is obtained as
\begin{equation}
c={ A + \sqrt{A^2+ 4\lambda_0 \lambda_1 (\lambda_{\rm av}-1)}
\over \lambda_0 \lambda_1} 
\label{eq:mfr}
\end{equation}
where $A=\lambda_0 \lambda_1-\lambda_0-\lambda_1$ and 
$\lambda_{\rm av}= P \lambda_1 + (1-P) \lambda_0$.
This mean-field approximation predicts that concentration vanishes
linearly when $\lambda_{\rm av} \to +1$. More precisely, for
low concentrations the above expression can be approximated as
\begin{equation}
c={P\lambda_1+(1-P)\lambda_0 - 1 \over
\lambda_0 + \lambda_1 - \lambda_0 \lambda_1}
\label{eq:mfa}
\end{equation}
if $\lambda_{\rm av} \geq 1$. Obviously, the trivial ($c=0$)
solution will be stable for $\lambda_{\rm av} < 1$.

In the limit $f \to \infty$ death and birth occur with
averaged rates: $\bar{d}=P d(1)+(1-P) d(0)$ and $\bar{b} =
P b(1) + (1-P) b(0)$), respectively. Thus the system becomes
equivalent to a homogeneous system (where $c_0=c_1=c$), and the
average concentration obeys the traditional form \cite{md99,h00}:
\begin{equation}
c={\bar{\lambda} - 1 \over \bar{\lambda}}
\label{eq:mfinfty}
\end{equation}
if $\bar{\lambda}=\bar{b} / \bar{d}\geq 1$ and $c=0$ otherwise.
The difference between the Eqs. (\ref{eq:mfr}) and (\ref{eq:mfinfty})
refers to the $f$ dependence. 

\section{SIMULATION OF SPREADING}
\label{sec:sim}

The Monte Carlo (MC) simulations are performed on a lattice
with $L \times L$ sites under periodic boundary conditions. 
Following the method suggested by Grassberger and de la Torre 
\cite{gdt} spreading is investigated by averaging over many 
trials $M$ when the initial state is close to the absorbing state.
More precisely, each run is started with a single individual [at
position ${\bf r}=(0,0)$] on a new, random, uncorrelated background.
We have determined the survival probability 
\begin{equation}
S(t)= \left\langle \theta \left( n(t) \right) \right\rangle \; ,
\label{eq:S}
\end{equation}
where the number of individuals at a given time $t$ is defined as
\begin{equation}
n(t)= \sum_{\bf r} \sigma({\bf r}) \; ,
\label{eq:nt}
\end{equation}
and $\langle \ldots \rangle$ means the average over $M$ trials
and $\theta (z)=1$ ($0$) for $z>0$ ($z\leq 0$).
We have also evaluated the average number of surviving individuals,
\begin{equation}
N(t)= \left\langle n(t) \right\rangle \; ,
\label{eq:N}
\end{equation}
and the mean-square distance of individuals from the origin,
\begin{equation}
R^2(t)={1 \over N(t)} \left\langle \sum_{\bf r} r^2 \sigma({\bf r})
\right\rangle \; .
\label{eq:R2}
\end{equation}

$M$ is varied from $10^4$ to $10^8$ in the systematic investigations.
The system size $L$ is adjusted to exceed significantly the average
radius of occupied sites, namely,
$L \agt 15 R(t_{max})$, where $t_{max}$ indicates the time limit.
As usual, time is measured in Monte Carlo Steps (MCS). During
one time unit, each site has a chance to modify the value of
$\sigma({\bf r})$ (approximately once on average). 

The efficiency of these simulations can be greatly improved by labeling
the individuals, recording their coordinates, and picking up one
individual randomly for updating the state of occupancy \cite{md99}.
In the present case, a chosen individual residing at site ${\bf r}$
dies with a probability $d(g({\bf r}))/(b(1)+d(0))$ or creates an
offspring on one of its nearest-neighboring sites ${\bf r}^{\prime}$
with a probability $b(g({\bf r}^{\prime}))/(b(1)+d(0))$. These attempted
events take time $\Delta t = 1/n$ on average, therefore time is
increased by $\Delta t$.
Variation of the background can be handled in the same way within
a smaller ($l \times l$) region whose sites affect the CP at the given
time. Namely, the values of $g({\bf r})$ are redrawn
at randomly chosen $\Delta t f l^2$ sites. 

The average values defined above [see Eqs. (\ref{eq:S})-(\ref{eq:R2})]
are determined for discrete time values chosen equidistantly in a
logarithmic scale.

\section{RESULTS FOR FROZEN BACKGROUND}
\label{sec:rfb}

\begin{figure}
\centerline{\epsfxsize=8cm
            \epsfbox{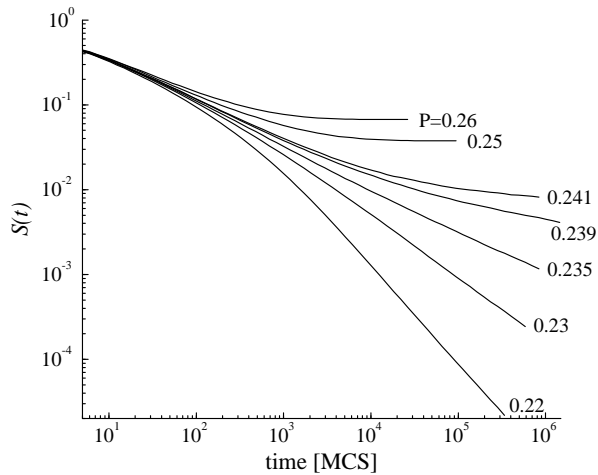}
            \vspace*{1mm}   }
\caption{Survival probabilities as functions of time at different $P$
values (indicated by labels) for $b(1)=1$, $b(0)=0.5$, $d(1)=0.25$,
and $d(0)=0.5$\ .}
\label{fig:stq}
\end{figure}

First we concentrate on a system where the random background is
frozen, i.e. $f=0$. Varying the proportion of good sites ($P$),
we have made a systematic MC analysis for fixed death and birth rates.
For the present parameters (see Fig.~\ref{fig:stq}) the good sites
form isolated clusters [having $P<P_{\rm perc}=0.5926$ (percolation
threshold)] and infinite spreading is possible across the bad sites.
The results summarized in the subsequent figures agree quantitatively
with those found by Moreira and Dickman \cite{md96} on a diluted
lattice for $P>P_{\rm perc}$.

\begin{figure}
\centerline{\epsfxsize=8cm
            \epsfbox{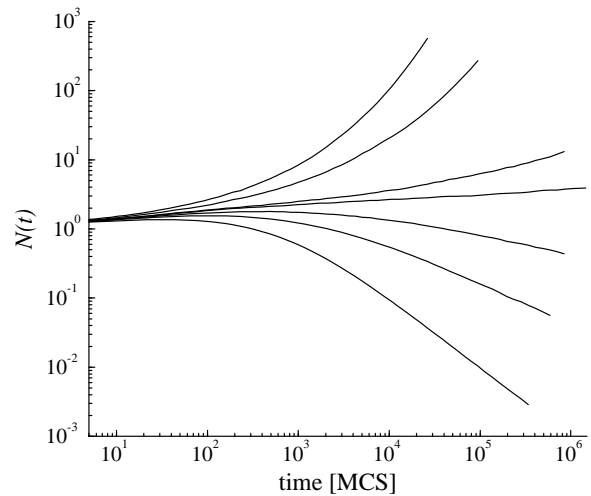}
            \vspace*{1mm}   }
\caption{Average number of individuals vs. time for the same simulations
as in Fig.~\ref{fig:stq}.}
\label{fig:ntq}
\end{figure}

Figure~\ref{fig:stq} shows some typical time-dependence functions
of survival in a log-log plot. The statistical errors are comparable
to the line width due to the large number of trials (e.g.
$M=10^8$ for $P=0.22$) . In the active region ($P>P_c$), $S(t)$ tends 
to a constant value as expected \cite{gdt}. In the subcritical region,
$S(t)$ can be well approximated by a power law. Unfortunately,
convergency toward the limit value becomes extremely slow in the
vicinity of the critical point, and this prevents the accurate
determination of $P_c$. Division between the active and inactive
regions becomes more visible when considering the function $N(t)$
(see Fig.~\ref{fig:ntq}).

In the active region, the surviving individuals occupy a compact
patch whose radius increases linearly with time. Consequently,
$N(t)$ becomes proportional to $t^2$ for sufficiently long
times as indicated in the Fig.~\ref{fig:ntq}. By contrast, when
$P<P_c$, then $N(t)$ tends to a power law decay with an exponent
depending on $P$.

\begin{figure}
\centerline{\epsfxsize=8cm
            \epsfbox{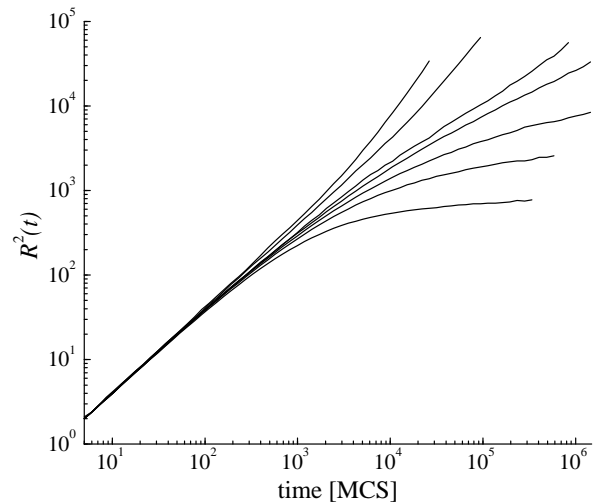}
            \vspace*{1mm}   }
\caption{Log-log plot of mean-square distance of individuals from
the origin as a function of time. Parameter values are the same
as in Fig.~\ref{fig:stq}.}
\label{fig:r2tq}
\end{figure}

Figure~\ref{fig:r2tq} shows the time evolution of the mean square
distance of surviving individuals as defined by Eq.~(\ref{eq:R2}). 
In agrement with the expectations $R^2(t) \propto t^2$ for sufficiently
long times if $P>P_c$. In the subcritical region, the increase of
$R^2(t)$ becomes significantly slower than that found for the
homogeneous system. In the homogeneous system, the surviving individuals
perform random walks independently of each other, therefore
$R^2(t) \propto t$ in the subcritical region. In the present case,
however, our data indicate a power law increase:
$R^2(t) \propto t^{\lambda}$, where the exponent $\lambda(P) < 1$. 
This indicates localization of spreading as, it will be explained
in Section \ref{sec:conc}.

\begin{figure}
\centerline{\epsfxsize=8cm
            \epsfbox{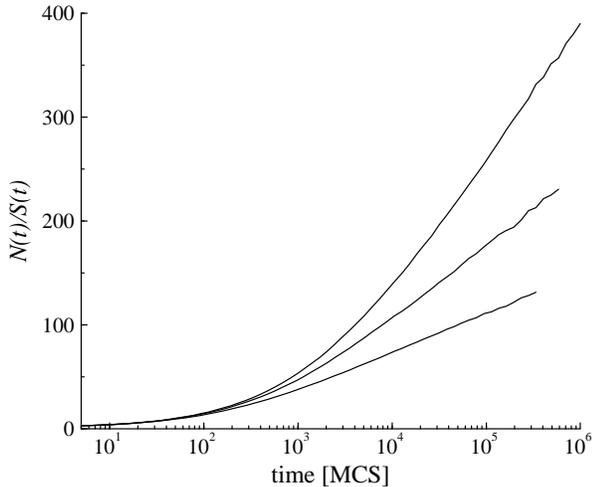}
            \vspace*{1mm}   }
\caption{Time dependence of the average number of individuals in the 
surviving trials at $P=0.22$, 0.23, and 0.235.}
\label{fig:npstq}
\end{figure}

Evaluation of $N(t)$ involves those (stoped) trials which
have reached the empty state before the given time $t$. Consequently,
the ratio $N(t)/S(t)$ expresses the average number of individuals
in the surviving trials. In the subcritical region, the time-dependence
of this quantity tends to logarithmic increase as
shown in a lin-log plot (see Fig.~\ref{fig:npstq}).

In the classical theory of percolation (reviewed in Refs. 
\cite{s79,e80}), a geometrical feature of the clusters of size
$s$ is characterized
by an average cluster radius $R_{\rm perc}(s)$. Below
the percolation threshold $R_{\rm perc}(s) \propto s^{\rho}$
for the large $s$ limit. According to MC simulations 
\cite{s79,leath,s78}, 
$\rho \simeq 2/3$, while the ``self-avoiding walk''
prediction \cite{s78z} gives $\rho \simeq 3/4$. In general,
$\rho > 1/2$
values are characteristic for ramified clusters \cite{s79}.
On the contrary, above the percolation threshold the finite
size clusters are compact, that yields $\rho = 1/2$.

\begin{figure}
\centerline{\epsfxsize=8cm
            \epsfbox{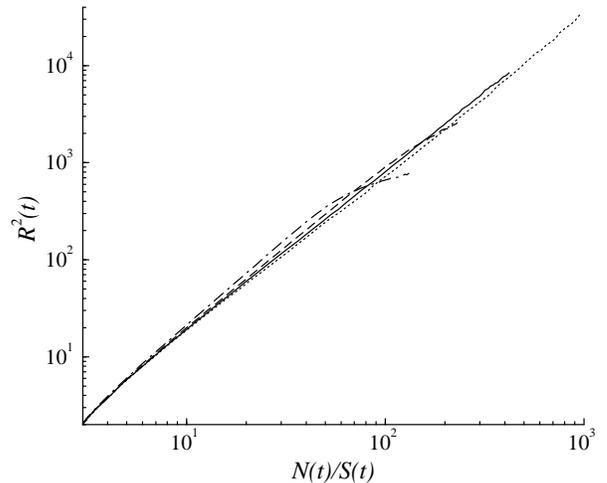}
            \vspace*{1mm}   }
\caption{$R^2(t)$ vs. $N(t)/S(t)$ for $P=0.22$ (dashed-dotted),
0.23 (dashed), 0.235 (solid), and 0.239 (dotted line).}
\label{fig:r2npsq}
\end{figure}

Following this analogy, we examine the relation between $N(t)/S(t)$,
the actual size of population in the surviving trials, and the
average radius of the area they occupy at a given time in the
subcritical region. Figure~\ref{fig:r2npsq} shows this relation
on a log-log plot. In the close vicinity of the critical point
(at $P=0.239$) we find power law behavior with an exponent 1.66(6).
This behavior indicates that the species spreads over a ramified
cluster in the whole time interval we could study the system.

In the subcritical region, however, we can observe two different
regions. In the transient (short-time) region, the increase of
$R^2(t)$ is faster than at the critical point. This indicates that
the offspring move away from the origin along preferred pathes
formed by good sites. Conversely, the long-time behavior
is dominated by those trials where the individuals stay on
a compact patch.

In the next section we show that these features fundamentally
change when we allow the background change over time.

\section{RESULTS FOR TIME-DEPENDENT BACKGROUND}
\label{sec:rtdb}

Applying time-dependent backgrounds, we determined the same
quantities [$S(t)$, $N(t)$, and $R^2(t)$] as before.
We use the same birth and death rates when varying
the value of $P$ for a fixed rate ($f$) of background change.
Our results estimate the critical value for the ratio of good sites
at $P_c=0.2680(1)$.

\begin{figure}
\centerline{\epsfxsize=8cm
            \epsfbox{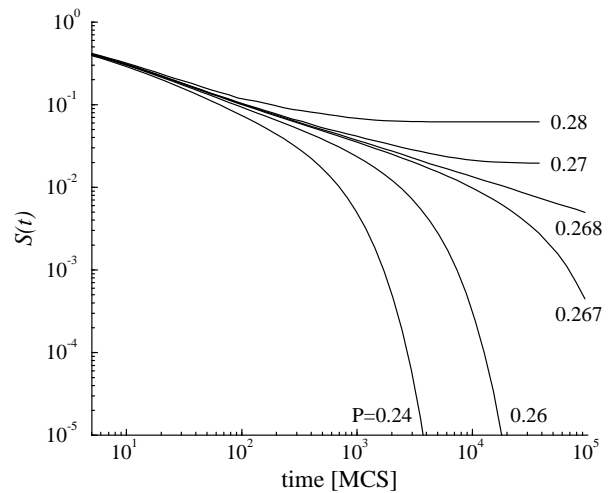}
            \vspace*{1mm}   }
\caption{Log-log plot of survival probabilities versus time at 
different $P$ values (as indicated by labels) for $b(1)=1$,
$b(0)=0.5$, $d(1)=0.25$, $d(0)=0.5$, and $f=0.01$\ .}
\label{fig:stt}
\end{figure}

Figure~\ref{fig:stt} shows that the survival probability
$S(t)$ vanishes exponentially when $P < P_c$. In the active
region ($P>P_c$), however, $S(t)$ tends to a constant value.
Notice, that the convergence is significantly faster here than
at the frozen background (see Fig.~\ref{fig:stq}).

Data for $P=0.268$ represent the behavior of the system at the
critical point. A numerical analysis confirms
that the survival probabilities can be well approximated by a 
power law, $S(t) \propto t^{-\delta}$ for sufficiently long times.
Numerical fitting gives $\delta = 0.45(2)$ in agreement with
the exponent found in DP transitions in homogeneous
systems \cite{md99,h00}.

\begin{figure}
\centerline{\epsfxsize=8cm
            \epsfbox{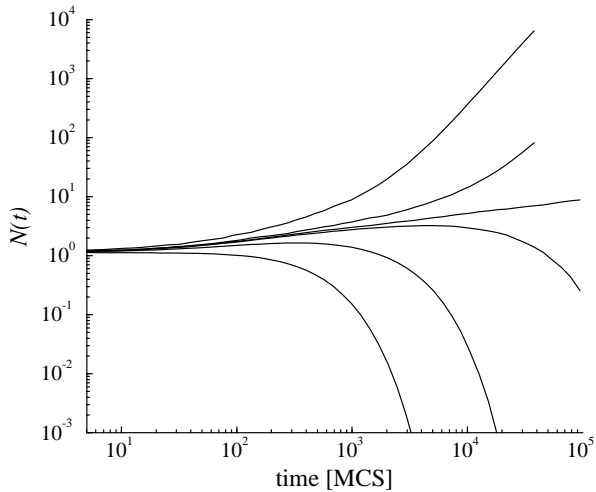}
            \vspace*{1mm}   }
\caption{Average number of individuals vs. time for the same
simulations as in Fig.~\ref{fig:stt}.}
\label{fig:ntt}
\end{figure}

When considering $N(t)$, we can also distinguish between three
different behaviors (see Fig.~\ref{fig:ntt}). Above the
critical point ($P>P_c$), $N(t)$ tends towards quadratic time
dependence. Below the critical point, $N(t)$ decreases exponentially.
At the critical point, however, our MC data indicate
power law behavior, i.e., $N(t) \propto t^{\eta}$
with $\eta = 0.23(1)$. This numerical value of $\eta$
agrees with the characteristic exponent of DP 
transitions \cite{md99,h00}.

In summary, all the above-mentiend numerical results support
that contact process exhibits the same features in time-dependent
random environments as in homogeneous background. Similarly to
$S(t)$ and $N(t)$, $R^2(t)$ also supports this conclusion (data
not shown).

These results have been gained at a fixed rate of changes
($f=0.01$). Preliminary results suggest, however, that the same
conclusions holds for other values of $f$ as well. General features
of spreading remain unchanged due to the robustness of DP transition.

Here it is worth mentioning that we have also studied the
concentration of individuals (as well as its fluctuation) in the
stationary state for some parameter values. These analyses support
that the variation of both quantities is consistent with the
expectations, e.i., the species concentration and its fluctuation
exhibit power law behavior with exponents characteristic for
the two-dimensional DP systems \cite{ad,bfm,md99,h00}.

\section{CONCLUSIONS AND SUMMARY}
\label{sec:conc}

We have studied a generalized version of contact process on a
square lattice. The background in the present model provides a
continuous transition from a homogeneous system to a randomly
fluctuating environment, including (frozen) diluted lattices.
The environment is inhomogeneous, consisting of two types
of sites (good and bad) that affect the rates of birth and
death of the species. In this model, spreading from a cluster
of good sites to another one is permitted across the bad sites
and/or bridges (of good sites) created occasionally by
background fluctuations.

Our analyses have been rescricted to the consideration of some
typical features, because the simulations are rather time-consuming.
The results confirm the theoretical expectations as well as the
general picture drawn by previous authors \cite{n86,n88,md96,dm98}
who considered CP on diluted lattices.

Obviously, an inhomogeneous environment contains areas that
provide better (or worse) conditions for the survival relative
to the average quality of the habitat. This variation becomes
particularly important in the vicinity of the transition point that
separates the active and inactive regions for long-time
behavior. In the subcritical region, the surviving individuals are
constrained to remain within isolated patches. Conversely, in the
active region, favorable areas can form a percolating cluster,
that has infinite extension, and sustains infinite survival.  

To explain the main results, we briefly remind the reader
of a simple calculation suggested by Noest \cite{n88}. In this
description we characterize the favorable patches by size $s$ and 
assume that their probability decreases exponentially,
i.e., $n_s \propto e^{-As}$. Furthermore, it is also assumed that
the average over many trials (or patches of size $s$) yields an
exponentially decreasing number of individuals over time, that is,
$m_s(t) \propto e^{-t/\tau_s}$ where the average survival time 
can be approximated as $\tau_s \propto e^{B s}$ if initially all
sites are occupied. Noest has shown that long-time behavior can
be well approximated in this case by a power law 
decrease, that is, $m(t)=\sum_s s n_s m_s(t) \propto t^{-A/B}$.
This prediction is derived from a maximum likelihood estimation
that implies the existence of a typical cluster
size giving the dominant contribution to $m(t)$. According to
this approach, the size of dominant clusters increases
logarithmically with time $t$. 

The adoption of this approach is not straightforward to the present
situation where we consider the spreading from a single seed.
In our case, the species can die out before achieving homogeneous
distribution, while homogeneous initial state is assumed in the
former calculation. We can assume, however, that the species spreads 
over the whole cluster or dies out within a transient time, and
the probabilities of these outcomes are independent of cluster
size. If this hypothesis holds, than the approach of Noest can be
applied for the analysis of long-time behavior.

Some predictions of this rough method have been confirmed by
our MC simulations. Namely, $S(t)$ and $N(t)$ exhibit a dominant
power law decrease in the subcritical region, while $N(t)/S(t)$
(in proportion to the dominant cluster size) increases logarithmically.
Furthermore, in the active region, the algebrically decaying 
contribution of $S(t)$ comes from extinction on (compact) finite size
clusters whose probability distribution satisfies the above
assumption \cite{s78,e80}.

Noest's approach is focused on compact clusters where
$s \propto R^2$. Consequently, $N(t)/S(t) \propto R^2(t)$ is
expected when the system behavior is dominated by a typical
cluster size. This behavior can be observed for long times
far below the critical point.
Figure~\ref{fig:r2npsq} indicates, however, that these predictions
(and assumptions) fail in the close vicinity of the critical point.
In other words, the ramified clusters give relevant contribution
to $S(t)$ and $N(t)$ during a transient time that increases when
approaching the critical point. In this case, we need a more
sophisticated description allowing the emergence of ramified 
clusters and the shape-dependence of $\tau_s$.

Numerical analysis of the extinction process is difficult (in a
quenched environment) because decrease of $S(t)$ and $N(t)$
slowes down in the vicinity of the critical point. In practice 
this implies persistence over long time periods before extinction.
The slow extinction process, however, requires rather specific
environmental conditions. Even a small degree of temporal fluctuation
results in an extinction process that is analogous to 
DP transition on large scales. Our mean-field calculations and
preliminary simulations have indicated that the transition
point is strongly affected by the rate $f$ of background change.
Further analyses are required to quantify this phenomenon for
different birth and death rates.

\acknowledgements
Support from the Hungarian National Research Fund (T-33098)
is acknowledged.

\end{document}